\documentclass[reprint,superscriptaddress,showpacs,amsmath,amssymb,aps,pra]{revtex4-2}
\usepackage{pifont}
\usepackage{times}
\usepackage{amssymb}
\usepackage{graphicx}
\usepackage{dcolumn}
\usepackage{dsfont}
\usepackage{bm}
\usepackage{subfigure}
\usepackage[ruled]{algorithm2e}
\usepackage{hyperref}
\usepackage{appendix}
\usepackage{lipsum}

\hypersetup{colorlinks=true, citecolor=blue, urlcolor=blue, linkcolor=blue}

\newtheorem{theorem}{Theorem}

\newtheorem{proposition}{Proposition}

\usepackage{amssymb}
\makeatletter
\newcommand{\rmnum}[1]{\romannumeral #1}
\newcommand{\Rmnum}[1]{\expandafter\@slowromancap\romannumeral #1@}

\newcommand{\ket}[1]{|#1 \rangle}
\newcommand{\braket}[2]{ \langle #1| #2 \rangle}

\input amssym.def

\begin{document}
\title{Quantifying the imaginarity via different distance measures}
\author{Meng-Li Guo}
\email{guomengli2020@163.com}
\affiliation{School of science, East China University of Technology, Nanchang 330006, China}
\author{Si-Yin Huang}
\email{h$_$siyin@126.com}
\affiliation{School of science, East China University of Technology, Nanchang 330006, China}
\author{Bo Li}
\email{libobeijing2008@163.com}
\affiliation{School of Computer and Computing Science, Hangzhou City University, Hangzhou 310015, China}
\author{Shao-Ming Fei}
%\email{feishm@cnu.edu.cn}
\affiliation{School of Mathematical Sciences, Capital Normal University, Beijing 100048, China}
	
\begin{abstract}
The recently introduced resource theory of imaginarity facilitates a systematic investigation into the role of complex numbers in quantum mechanics and quantum information theory. In this work, we propose well-defined measures of imaginarity using various distance metrics, drawing inspiration from recent advancements in quantum entanglement and coherence. Specifically, we focus on quantitatively evaluating imaginarity through measures such as Tsallis relative $\alpha$-entropy, Sandwiched R\'{e}nyi relative entropy, and Tsallis relative operator entropy. Additionally, we analyze the decay rates of these measures. Our findings reveal that the Tsallis relative $\alpha$-entropy of imaginarity exhibits higher decay rate under quantum channels compared to other measures. Finally, we examine the ordering of single-qubit states under these imaginarity measures, demonstrating that the order remains invariant under the bit-flip channel for specific parameter ranges. This study enhances our understanding of imaginarity as a quantum resource and its potential applications in quantum information theory.
\end{abstract}
	
\maketitle
	
\section{Introduction}
Complex numbers form a cornerstone in the formulation of various physical theories, notably in mechanics, electrodynamics, and optics, where they provide elegant and concise descriptions of physical phenomena. In quantum mechanics, however, the role of imaginary numbers extends beyond simplification; they are intrinsic to the theory's structure. Historically, imaginary numbers have been used to model oscillatory motion and wave mechanics in classical physics, but their role in quantum physics is fundamentally deeper. They are essential to any standard formulation of quantum mechanics, as evidenced by numerous foundational studies \cite{Wootters37,Hardy38,Aleksandrova39}.

Consider, for example, the polarization density matrix of a single photon. In the $\{|H\rangle, |V\rangle\}$ basis, where $|H\rangle$ and $|V\rangle$ represent horizontal and vertical polarization, respectively, the presence of imaginary numbers in the density matrix induces rotation of the electric field vector, resulting in elliptical or circular polarization. This example highlights the crucial role of imaginary numbers in describing fundamental properties of quantum systems.

The question arises: is it necessary to invoke the imaginary part of the wave function $\Psi(x) = |\Psi(x)|e^{-i\phi(x)}$ to describe the dynamics of a quantum system? Specifically, can quantum mechanics be reformulated using only real numbers? Recent developments in quantum resource theory provide a definitive answer. While various real-number formulations of quantum mechanics have been proposed, they cannot fully capture the richness and complexity of quantum phenomena \cite{Wu1,Renou2,Chen3,Li4}. Notably, certain experimental scenarios cannot be accurately modeled using quantum theory with real-valued amplitudes alone, underscoring the essential role of imaginary numbers in quantum mechanics.

Quantum resource theory provides a unified framework for studying various quantum phenomena and their applications in quantum information protocols. It offers a rigorous approach to quantify and manipulate resources within quantum systems, including entanglement \cite{Horodecki1}, coherence \cite{Baumgratz4,Chitambar52,Marvian6,ChitambarHsieh7,Winter8,Napoli9}, athermality \cite{Brandao10,Horodecki11,Lostaglio12,Gour13,Narasimhachar14}, asymmetry \cite{Gour15,Gour16,Skotiniotis17,Marvian18}, knowledge \cite{Rio19}, magic \cite{Veitch20,Howard21,Ahmadi22}, superposition \cite{Theurer23}, steering \cite{Gallego33}, nonlocality \cite{Vicente34}, and contextuality \cite{Amaral35}. Within this framework, free states and free operations are defined, facilitating the identification and quantification of resource states.

Recently, the concept of imaginarity as an operational resource has been introduced \cite{Hickey1}. Building on recent advances in entanglement and coherence theories, Wu et al. developed a comprehensive theory of imaginarity resources \cite{Wu1,Wu17}. They quantified imaginarity using specific measures, such as geometric imaginarity and robustness of imaginarity, and provided operational interpretations of these measures in state transformation contexts. Their findings illustrate that imaginarity serves as a valuable resource in optical experiments and deepens the fundamental understanding of imaginarity resource theory.

A quantitative analysis of imaginarity not only deepens our understanding of this resource but also advances the development of quantum technologies, such as quantum computing. Different methods for quantifying imaginarity offer varied insights into its properties and applications. Several specific measures of imaginarity have been proposed and studied, including the trace norm of imaginarity \cite{Hickey1,Wu1}, the fidelity of imaginarity \cite{Wu17,Kondra11}, the relative entropy of imaginarity \cite{Xue7}, geometric-like imaginarity \cite{Guo2024}, the weight of imaginarity \cite{Xue7}, and the imaginarity of bosonic Gaussian states \cite{Xu1}. Nevertheless, further research is required to fully explore the potential of imaginarity as a resource in quantum mechanics and its applications.

In this paper, within the framework of resource theory, we introduce imaginarity measures based on the Tsallis relative $\alpha$-entropy, Sandwiched R\'{e}nyi relative entropy, and Tsallis relative operator entropy. The remainder of the paper is organized as follows. In Section 2, we review foundational concepts, including real states, free operations, and measures of imaginarity. Subsequently, we investigate the imaginarity measures based on the Tsallis relative $\alpha$-entropy, Sandwiched R\'{e}nyi relative entropy, and Tsallis relative operator entropy within the resource theory framework. In Section 3, we analyze the decay rates of these measures. In Section 4, we discuss the ordering of qubit states and the influence of quantum channels on the state ordering of a single qubit.

\section{Imaginarity measures based on different distances}
Let $\{|j\rangle \}^{d-1} _ {j = 0} $ be a fixed base in $d$-dimensional Hilbert space $\mathcal{H}$. We denote $\mathcal{D}(\mathcal{H})$ the set of density operators acting on $\mathcal{H}$. In the resource theory of imaginarity, the free states are real states given by
\begin{eqnarray}
	\mathcal{F}=\{\rho \in \mathcal{D}(\mathcal{H}): \langle m|\rho|n \rangle \in \mathbb{R}, m,n=0,1,...,d-1\},\nonumber
\end{eqnarray}
where $\mathbb{R}$ denotes the set of real numbers. The real operations are those that can be characterized by real Kraus operators. A operation $\varepsilon$ characterized by Kraus operators $\{K_j\}$ is real if $\langle m|K_j|n \rangle \in \mathbb{R}$ for any $j$ and $m,n=0,1,...,d-1$. A real-valued functional $M$ of quantum states is called an imaginarity measure if it satisfies the following conditions (M1) to (M4) \cite{Wu1,Wu17,Hickey1}.

(M1) $Nonnegativity$ $M(\rho )\geq 0$ and $M(\rho )=0$ if and only if $\rho$ is a real state.

(M2) $Monotonicity$ $M[\varepsilon(\rho)]\leq M(\rho)$ if $\varepsilon$ is a real operation.

(M3) $Strong~imaginarity~monotonicity$ $\sum_j p_j M(\rho_j)\leq M(\rho)$, where the probabilities $p_j=\mathrm{Tr}K_j\rho K^{\dagger}_j$, $\rho_j= K_j\rho K^{\dagger}_j/p_j$ and $K_j$ are real Kraus operators.

(M4) $Convexity$ $M(\sum_{j}p_{j}\rho _{j})\leq \sum_{j}p_{j}M(\rho _{j})$ for any probability distribution $\{p_{j}\}$ and states $\{\rho_{j}\}.$

In \cite{Xue7} the authors prove that (M1) to (M4) are equivalent to (M1), (M2) and (M5),

(M5) $Additivity$ $M[p\rho_{1}\oplus (1-p)\rho_{2}]=pM(\rho_{1})+(1-p)M(\rho_{2})$, where $p\in (0,1),$ $\{\rho_{1},\rho_{2}\}$ are any quantum states.

For $\alpha \in \lbrack 0,1)$, the Tsallis relative entropy of states $\rho$ and $\sigma$ is defined as \cite{Borland21,Furuichi22,Abe23}
\begin{eqnarray}
	D_{\alpha}(\rho ||\sigma )=\frac{1-\text{tr}(\rho^{\alpha}\sigma^{1-\alpha})}{1-\alpha}.
\end{eqnarray}
For $\alpha \in \lbrack \frac{1}{2},1)$, the sandwiched R\'{e}nyi relative entropy is defined as \cite{Wilde36,Muller37}
\begin{eqnarray}
	F_{\alpha }(\sigma ||\rho )=\frac{\ln \text{tr}[(\rho ^{\frac{1-\alpha }{2\alpha }}\sigma \rho ^{\frac{1-\alpha }{2\alpha }})^{\alpha }]}{\alpha -1}. \nonumber
\end{eqnarray}
For $\alpha\in [0,1)$, the Tsallis relative operator entropy is defined by
\begin{equation}\label{n2}
	T_{\alpha}(\rho||\sigma)=\frac{\rho^{\frac{1}{2}}(\rho^{-\frac{1}{2}}\sigma \rho^{-\frac{1}{2}})^{1-\alpha}\rho^{\frac{1}{2}}-\rho}{\alpha-1}.
\end{equation}

Below we propose several quantum imaginarity measures based on Tsallis relative entropy, sandwiched R\'{e}nyi relative entropy and Tsallis relative operator entropy. We only consider the case of $\alpha \in [\frac{1}{2},1)$ in this article. The proof of theorem \ref{main1} is given in Appendix.
\begin{theorem}	\label{main1}
	For $\alpha \in [\frac{1}{2},1)$, $\rho \in \mathcal{D}(\mathcal{H})$, all the following quantities given in (\rmnum{1}) to (\rmnum{3}) are bona fide measures of quantum imaginarity.
	
	(\rmnum{1}) Imaginarity based on Tsallis relative $\alpha$-entropy
	\begin{equation}\label{Ta}
		\mathcal{M}_{T,\alpha}(\rho)=\min_{\sigma \in \mathcal{F}} \Bigg\{1-\Big[\text{tr}(\rho^{\alpha}\sigma^{1-\alpha})\Big]^{\frac{1}{\alpha}}\Bigg\}.\nonumber
	\end{equation}
	
	(\rmnum{2}) Sandwiched R\'{e}nyi relative entropy of imaginarity
	\begin{equation}\label{Ra}
		\mathcal{M}_{S,\alpha}(\rho)=\min_{\sigma \in \mathcal{F}}\frac{1}{\alpha-1}\Bigg\{\Big\{\text{tr}\Big[(\rho^{\frac{1-\alpha}{2\alpha}}\sigma\rho^{\frac{1-\alpha}{2\alpha}})^{\alpha}\Big]\Big\}^{\frac{1}{1-\alpha}}-1\Bigg\}.\nonumber
	\end{equation}
	
	(\rmnum{3}) Tsallis relative operator entropy of imaginarity
	\begin{eqnarray}\label{Oa}
		&&\mathcal{M}_{O,\alpha}(\rho)\nonumber\\
		&=&\min_{\sigma \in \mathcal{F}}\frac{1}{\alpha-1} \Bigg\{\Big[\text{tr}\Big(\rho^{\frac{1}{2}}(\rho^{-\frac{1}{2}}\sigma \rho^{-\frac{1}{2}})^{1-\alpha}\rho^{\frac{1}{2}}\Big)\Big]^{\frac{1}{\alpha}}-1\Bigg\}.\nonumber
	\end{eqnarray}
\end{theorem}

In particular, for any pure states $\ket{\psi}$, the imaginarity is just a function of $|\braket{\psi^*}{\psi}|$.
\begin{theorem}\label{mian3}
	For qubit pure state $\ket{\psi}$, we have
	\begin{eqnarray}
		\mathcal{M}_{T,\alpha}(\ket{\psi})&=&\Big[1-\Big(\frac{1
			+|\braket{\psi^*}{\psi}|}{2}\Big)^{\frac{1}{\alpha} }\Big],\nonumber\\
		\mathcal{M}_{S,\alpha}(\ket{\psi})&=&\frac{1}{\alpha -1} \Big[\Big(\frac{1+|\braket{\psi^*}{\psi}|}{2}\Big)^{\frac{\alpha }{1-\alpha }}-1\Big],\nonumber\\
		\mathcal{M}_{O,\alpha}(\ket{\psi})&=&\frac{1}{\alpha -1}\Big[\Big(\frac{1+|\braket{\psi^*}{\psi}|}{2}\Big)^{\frac{1-\alpha }{\alpha }}-1\Big].\nonumber
	\end{eqnarray}
\end{theorem}	

{\sf [Proof]} 
For any pure state $\ket{\psi}$, there exists a real orthogonal matrix $O$ such that
\begin{equation}
	O\ket{\psi} =\sqrt{\frac{1+|\braket{\psi^*}{\psi}|}{2}}\ket{0} + i \sqrt{\frac{1-|\braket{\psi^*}{\psi}|}{2}}\ket{1},\label{opsi}
\end{equation}
which keeps the imaginarity unchanged \cite{Wu17}.
To complete the proof, we now evaluate $\mathcal{M}$ for any state of
the form $\ket{\psi} = a_0\ket{0} + ia_1\ket{1}$ be a general qubit state with $a_0^2 + a_1^2 = 1$ and $ a_0^2\geq a_1^2\geq 0$. For any real state $\ket{v} =b_0\ket{0}+b_1\ket{1}$ with $b_0^2 + b_1^2=1$, we have
\begin{eqnarray}
	\mathcal{M}_{T,\alpha}(\ket{\psi})&=&\min_{b_0,b_1} \Big[1-(2a_0^2b_0^2-a_0^2-b_0^2+1)^{\frac{1}{\alpha} }\Big],\nonumber\\
	\mathcal{M}_{S,\alpha}(\ket{\psi})&=&\min_{b_0,b_1}\frac{1}{\alpha -1} \Big[(2a_0^2b_0^2-a_0^2-b_0^2+1)^{\frac{\alpha }{1-\alpha }}-1\Big],\nonumber\\
	\mathcal{M}_{O,\alpha}(\ket{\psi})&=&\min_{b_0,b_1}\frac{1}{\alpha -1}\Big[(2a_0^2b_0^2-a_0^2-b_0^2+1)^{\frac{1-\alpha }{\alpha }}-1\Big].\nonumber
\end{eqnarray}
Since $0\leq2a_0^2b_0^2-a_0^2-b_0^2+1 \leq a_0^2,$ we obtain $(2a_0^2b_0^2-a_0^2-b_0^2+1)^n \leq (a_0^2)^n$ for $n=\frac{1}{\alpha}$, ${\frac{\alpha }{1-\alpha }}$ and ${\frac{1-\alpha }{\alpha }}$. Therefore, $\max_{b_0,b_1}(2a_0^2b_0^2-a_0^2-b_0^2+1)^n=(a_0^2)^n$ and
\begin{eqnarray}\label{3p}
	\mathcal{M}_{T,\alpha}(\ket{\psi})&=&\Big[1-(a_0^2)^{\frac{1}{\alpha} }\Big],\nonumber\\
	\mathcal{M}_{S,\alpha}(\ket{\psi})&=&\frac{1}{\alpha -1} \Big[(a_0^2)^{\frac{\alpha }{1-\alpha }}-1\Big],\nonumber\\
	\mathcal{M}_{O,\alpha}(\ket{\psi})&=&\frac{1}{\alpha -1}\Big[(a_0^2)^{\frac{1-\alpha }{\alpha }}-1\Big].
\end{eqnarray}
From (\ref{3p}) and (\ref{opsi}) we prove the Theorem \ref{mian3}.
$\hfill\blacksquare$

Later we will consider the relationships among the Tsallis relative $\alpha$-entropy of imaginarity, Sandwiched R\'{e}nyi relative entropy of imaginarity and Tsallis relative operator entropy of imaginarity. For pure states we have the following result.
\begin{proposition}	
	For any pure qubit state $\ket{\psi}$, we have
	\begin{eqnarray*}\label{PureRe}
		\mathcal{M}_{S,\alpha}(\ket{\psi})\geq \mathcal{M}_{O,\alpha}(\ket{\psi}) \geq \mathcal{M}_{T,\alpha}(\ket{\psi}),
	\end{eqnarray*}
	where $\alpha \in [\frac{1}{2},1)$.
\end{proposition}

{\sf [Proof]} Denote $A=|\braket{\psi^*}{\psi}|$. We have
\begin{eqnarray}
	\mathcal{M}_{T,\alpha}(\ket{\psi})&=&\Big[1-\Big(\frac{1+A}{2}\Big)^{\frac{1}{\alpha} }\Big],\nonumber\\
	\mathcal{M}_{S,\alpha}(\ket{\psi})&=&\frac{1}{\alpha -1} \Big[\Big(\frac{1+A}{2}\Big)^{\frac{\alpha }{1-\alpha }}-1\Big],\nonumber\\
	\mathcal{M}_{O,\alpha}(\ket{\psi})&=&\frac{1}{\alpha -1}\Big[\Big(\frac{1+A}{2}\Big)^{\frac{1-\alpha }{\alpha }}-1\Big].\nonumber
\end{eqnarray}
Let $\Delta M_1=\mathcal{M}_{S,\alpha}(\ket{\psi})-\mathcal{M}_{O,\alpha}(\ket{\psi})$, then
\begin{eqnarray}
	\Delta M_1=\frac{1}{\alpha -1} \Big[\Big(\frac{1+A}{2}\Big)^{\frac{\alpha }{1-\alpha }}-\Big(\frac{1+A}{2}\Big)^{\frac{1-\alpha }{\alpha }}\Big].\nonumber
\end{eqnarray}
Since $\alpha \in [\frac{1}{2},1)$ and $A\in[0,1]$, we have $\frac{1}{\alpha -1}<0$, $\frac{\alpha}{1-\alpha}\geq \frac{1-\alpha}{\alpha}\geq0$, $\frac{1+A}{2}\in[\frac{1}{2},1]$.
When $0<a<1$, the function $F(x)=a^x$ exhibits a monotonically decreasing property within its domain, thus it follows that
\begin{eqnarray}
	&&\Big(\frac{1+A}{2}\Big)^{\frac{\alpha }{1-\alpha }}\leq \Big(\frac{1+A}{2}\Big)^{\frac{1-\alpha }{\alpha }}.\nonumber
\end{eqnarray}
Therefore, it is verified that $\Delta M_1 \geq0$, namely, $\mathcal{M}_{S,\alpha}(\ket{\psi})\geq \mathcal{M}_{O,\alpha}(\ket{\psi})$.

\begin{figure}
	\includegraphics[width=9cm]{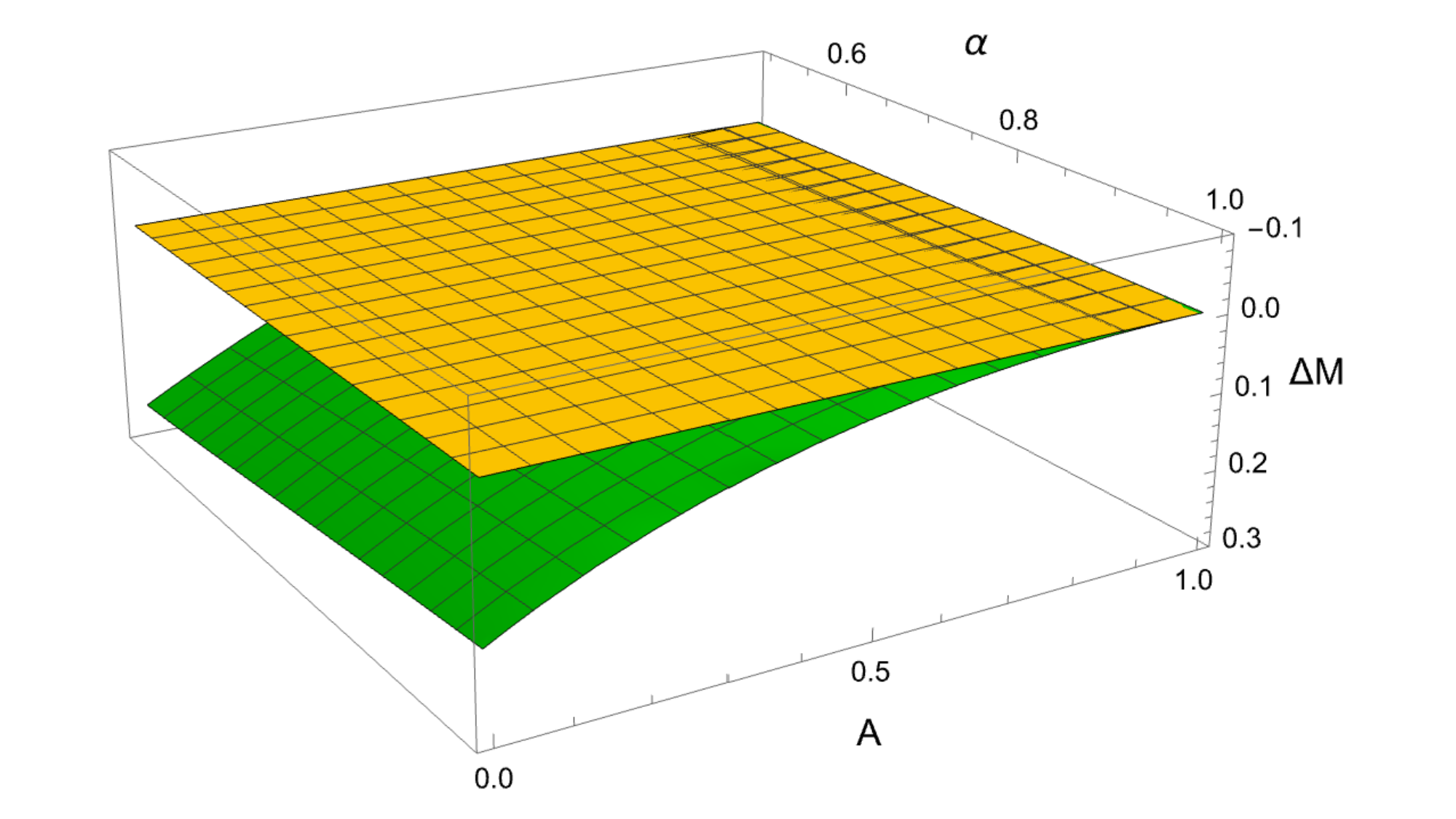}
	\caption{The green surface represents $\Delta M_2$, and the yellow surface represents 0.}
	\label{IM1}
\end{figure}

Similarly, let
$\Delta M_2=\mathcal{M}_{O,\alpha}(\ket{\psi})-\mathcal{M}_{T,\alpha}(\ket{\psi})$,
we have 
\begin{eqnarray}
	\Delta M_2&=&\frac{1}{\alpha -1} \Big[\Big(\frac{1+A}{2}\Big)^{\frac{1-\alpha }{\alpha }}-\Big(\frac{1+A}{2}\Big)^{\frac{1}{\alpha}}\nonumber\\
	&&+\alpha \Big(\frac{1+A}{2}\Big)^{\frac{1}{\alpha}}-\alpha \Big]\nonumber\\
	&=&\frac{2^{\frac{1}{\alpha }} (A+1)^{\frac{1}{\alpha}-1}[\alpha +(\alpha -1) A+1]-\alpha }{\alpha -1}.\nonumber
\end{eqnarray}
As can be seen from Fig. \ref{IM1}, when $A\in[0,1]$ and $\alpha \in [\frac{1}{2},1)$,
$\Delta M_2 \geq0$, which implies $\mathcal{M}_{O,\alpha}(\ket{\psi}) \geq \mathcal{M}_{T,\alpha}(\ket{\psi})$.
$\hfill\blacksquare$

\section{Decay of imaginarity of pure states under quantum channels}\label{sec:4}
Quantum systems are inherently influenced by their environments. Similar to quantum entanglement and coherence, the imaginarity of a quantum state is also affected by environmental factors. Analyzing the impact of noisy environments on the imaginarity of quantum states is essential. This section aims to investigate the Tsallis relative $\alpha$-entropy of imaginarity, the sandwiched R\'{e}nyi relative entropy of imaginarity, and the Tsallis relative operator entropy of imaginarity for pure states under various quantum channels. Specifically, we explore how quantum noise affects these measures of imaginarity, drawing parallels to studies on coherence decay \cite{Xi56}.

For a general quantum channel $\varepsilon$, the decay of imaginarity is defined as the difference between the initial state's imaginarity and that after the application of the noisy channel, i.e., $\Delta\mathcal{M} = \mathcal{M}(\rho) - \mathcal{M}(\varepsilon(\rho))$. According to (\ref{opsi}) and the fact that the imaginarity of any pure state $\rho$ remains unchanged under a real orthogonal transformation $O$, we have $\mathcal{M}(O\rho O^T) = \mathcal{M}(\rho)$. We consider the following general form \cite{Guo2024}:
\begin{eqnarray}
	\ket{\eta}&=&\sqrt{\frac{1+|\braket{\psi^*}{\psi}|}{2}}\ket{0} + i \sqrt{\frac{1-|\braket{\psi^*}{\psi}|}{2}}\ket{1}\nonumber\\
	&=&\sqrt{\frac{1+A}{2}}\ket{0} + i \sqrt{\frac{1-A}{2}}\ket{1},\nonumber
\end{eqnarray}
where $A=|\braket{\psi^*}{\psi}|$.

Consider the bit flip (BF), phase damping (PD) and amplitude damping (AD) channels given by the following Kraus operators, respectively,
\begin{align}
	E^{BF}_0&=\left(
	\begin{array}{cc}
		\sqrt{m} & 0 \\
		0 & \sqrt{m} \\
	\end{array}
	\right),~~E^{BF}_1= \left(
	\begin{array}{cc}
		0 & \sqrt{1-m} \\
		\sqrt{1-m} & 0 \\
	\end{array}
	\right);\nonumber\\
	E^{PD}_0&=\left(
	\begin{array}{cc}
		1 & 0 \\
		0 & \sqrt{1-n} \\
	\end{array}
	\right),~~E^{PD}_1=\left(
	\begin{array}{cc}
		0 & 0 \\
		0 & \sqrt{n} \\
	\end{array}
	\right);\nonumber\\
	E^{AD}_0&=\left(
	\begin{array}{cc}
		1 & 0 \\
		0 & \sqrt{1-p} \\
	\end{array}
	\right),~~E^{AD}_1=\left(
	\begin{array}{cc}
		0 & \sqrt{p} \\
		0 & 0 \\
	\end{array}
	\right).\nonumber
\end{align}
Under the BF, PD and AD channels, a state $\rho$ is transformed into, respectively,
\begin{eqnarray}
	\rho_{BF}=\left(
	\begin{array}{cc}
		A \left(m-\frac{1}{2}\right)+\frac{1}{2} & \frac{1}{2} i \sqrt{1-A^2} (1-2 m) \\
		\frac{1}{2} i \sqrt{1-A^2} (2 m-1) & \frac{1}{2} (-2 A m+A+1) \\
	\end{array}
	\right),\nonumber
\end{eqnarray}
\begin{eqnarray}
	\rho_{PD}=\left(
	\begin{array}{cc}
		\frac{A+1}{2} & -\frac{1}{2} i \sqrt{1-A^2} \sqrt{1-n} \\
		\frac{1}{2} i \sqrt{1-A^2} \sqrt{1-n} & \frac{1-A}{2} \\
	\end{array}
	\right),\nonumber
\end{eqnarray}
\begin{eqnarray}
	\rho_{AD}=\left(
	\begin{array}{cc}
		\frac{1}{2} (A (-p)+A+p+1) & -\frac{1}{2} i \sqrt{1-A^2} \sqrt{1-p} \\
		\frac{1}{2} i \sqrt{1-A^2} \sqrt{1-p} & \frac{1}{2} (A-1) (p-1) \\
	\end{array}
	\right).\nonumber
\end{eqnarray}
Through meticulous algebraic manipulations, we obtain the following expressions for the decay of the Tsallis relative $\alpha$-entropy of imaginarity, Sandwiched R\'{e}nyi relative entropy of imaginarity and Tsallis relative operator entropy of imaginarity under these channels with $\alpha=\frac{3}{4}$,
\begin{eqnarray}
	\Delta\mathcal{M}^{BF}_{T,\frac{3}{4}}&=&\frac{1}{4} \Big[2^{\frac{2}{3}} \Big(\frac{\sqrt{s_1-t_1}+\sqrt{s_1+t_1}}{\sqrt{2}}+1\Big)^{\frac{4}{3}}\nonumber\\
	&&-2^{\frac{2}{3}} (A+1)^{\frac{4}{3}}\Big],\nonumber\\
	\Delta\mathcal{M}^{PD}_{T,\frac{3}{4}}&=&\frac{1}{8} \Big[\sqrt[3]{2} \Big(\sqrt{s_2-t_2}+\sqrt{s_2+t_2}+2\Big)^{\frac{4}{3}}\nonumber\\
	&&-2^{\frac{5}{3}} (A+1)^{\frac{4}{3}}\Big],\nonumber\\
	\Delta\mathcal{M}^{AD}_{T,\frac{3}{4}}&=&\frac{(\sqrt{s_3-t_3}+\sqrt{s_3+t_3}+2)^{\frac{4}{3}}-2 \sqrt[3]{2} (A+1)^{\frac{4}{3}}}{ 2^{\frac{8}{3}}};\nonumber
\end{eqnarray}
\begin{eqnarray}
	\Delta\mathcal{M}^{BF}_{S,\frac{3}{4}}&=&\frac{1}{2} \Big(\frac{\sqrt{s_1-t_1}+\sqrt{s_1+t_1}}{\sqrt{2}}+1\Big)^3-\frac{1}{2}(A+1)^3,\nonumber\\
	\Delta\mathcal{M}^{PD}_{S,\frac{3}{4}}&=&\frac{1}{16} \Big(\sqrt{s_2-t_2}+\sqrt{s_2+t_2}+2\Big)^3-\frac{1}{2}(A+1)^3,\nonumber\\
	\Delta\mathcal{M}^{AD}_{S,\frac{3}{4}}&=&\frac{1}{16} \Big(\sqrt{s_3-t_3}+\sqrt{s_3+t_3}+2\Big)^3-\frac{1}{2}(A+1)^3,\nonumber
\end{eqnarray}
and
\begin{eqnarray}
	\Delta\mathcal{M}^{BF}_{O,\frac{3}{4}}&=&2^{\frac{4}{3}}\sqrt[3]{\sqrt{2} \sqrt{s_1-t_1}+\sqrt{2} \sqrt{s_1+t_1}+2}\nonumber\\
	&&-2^{\frac{5}{3}} \sqrt[3]{A+1},\nonumber\\
	\Delta\mathcal{M}^{PD}_{O,\frac{3}{4}}&=&2^{\frac{4}{3}} \sqrt[3]{\sqrt{s_2-t_2}+\sqrt{s_2+t_2}+2}-2^{\frac{5}{3}}\sqrt[3]{A+1},\nonumber\\
	\Delta\mathcal{M}^{AD}_{O,\frac{3}{4}}&=&2^{\frac{4}{3}} \sqrt[3]{\sqrt{s_3-t_3}+\sqrt{s_3+t_3}+2}-2^{\frac{5}{3}}\sqrt[3]{A+1},\nonumber
\end{eqnarray}
where $s_1=A^2 (1-2 m)^2-2 (m-1) m$, $t_1=A (1-2 m) \sqrt{A^2 (1-2 m)^2-4 (m-1) m}$, $s_2=n-A^2 (n-2)$, $t_2=2 \sqrt{A^4 (-n)+A^2 n+A^4}$, $s_3=A^2 (p-2) (p-1)-2 A (p-1) p+p^2+p$ and $t_3=2 \sqrt{\left(p-A^2 (p-1)\right) (A (-p)+A+p)^2}$.

\begin{widetext}
	\begin{figure*}[tb]
		\includegraphics[width=18cm]{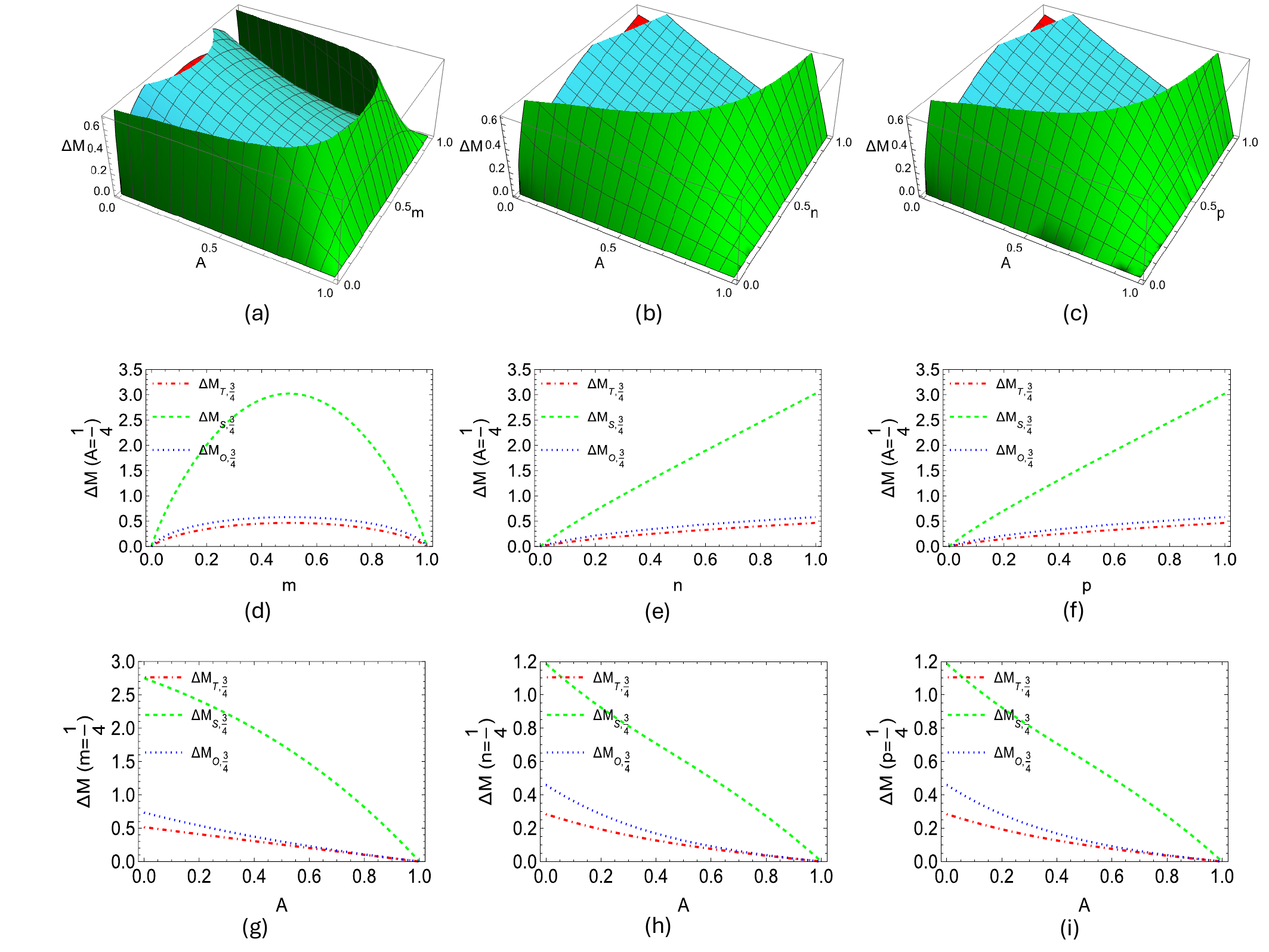}
		\caption{The decay of $\Delta\mathcal{M}_{T,\frac{3}{4}}$, $\Delta\mathcal{M}_{S,\frac{3}{4}}$, and $\Delta\mathcal{M}_{O,\frac{3}{4}}$ with respect to different channels is illustrated in Fig. \ref{Fig:1}. Panels (a), (d), and (g) depict the attenuation for the bit flip channel, corresponding to $\Delta\mathcal{M}_{gl}$ and $\Delta\mathcal{M}_{g}$. Panels (b), (e), and (h) represent the phase damping channel's attenuation for $\Delta\mathcal{M}_{T,\frac{3}{4}}$, $\Delta\mathcal{M}_{S,\frac{3}{4}}$, and $\Delta\mathcal{M}_{O,\frac{3}{4}}$. Lastly, panels (c), (f), and (i) illustrate the amplitude damping channel attenuation for the same decay functions. The red surface in panels (a), (b), and (c) denotes $\Delta\mathcal{M}_{T,\frac{3}{4}}$, while the green surface represents $\Delta\mathcal{M}_{S,\frac{3}{4}}$, and the blue surface corresponds to $\Delta\mathcal{M}_{O,\frac{3}{4}}$.}
		\label{Fig:1}
	\end{figure*}
\end{widetext}

According to the above equations, the decay functions $\Delta\mathcal{M}_{T,\frac{3}{4}}$, $\Delta\mathcal{M}_{S,\frac{3}{4}}$, and $\Delta\mathcal{M}_{O,\frac{3}{4}}$ are illustrated in Fig. \ref{Fig:1} for various scenarios under fixed quantum states (fixed $A$) and fixed channels (fixed $m$, $n$, $p$).
As shown in Figs. 2(a) and 2(d), for any bit flip channel, $\Delta\mathcal{M}_{T,\frac{3}{4}}$, $\Delta\mathcal{M}_{S,\frac{3}{4}}$, and $\Delta\mathcal{M}_{O,\frac{3}{4}}$ all reach their maximum values at $m=0.5$, and are zero at $m=0$ or $m=1$. However, as depicted in Figs. 2(b), 2(e), 2(c), and 2(f), for phase flip and amplitude damping channels, these quantities do not attain their maximum values at $m=0.5$.
According to Figs. 2(d) and 2(g), for a fixed quantum state (fixed $A$), $\Delta\mathcal{M}_{T,\frac{3}{4}}$, $\Delta\mathcal{M}_{S,\frac{3}{4}}$, and $\Delta\mathcal{M}_{O,\frac{3}{4}}$ exhibit concave behavior with respect to $m$ (different bit flip channels); conversely, for a fixed bit flip channel (fixed $m$), $\Delta\mathcal{M}_{S,\frac{3}{4}}$ exhibits concave behavior with respect to $A$ (different quantum states), while $\Delta\mathcal{M}_{T,\frac{3}{4}}$ and $\Delta\mathcal{M}_{O,\frac{3}{4}}$ exhibit convex behavior.
Similarly, according to Figs. 2(e) and 2(h), when the quantum state is fixed (fixed $A$), $\Delta\mathcal{M}_{T,\frac{3}{4}}$, $\Delta\mathcal{M}_{S,\frac{3}{4}}$, and $\Delta\mathcal{M}_{O,\frac{3}{4}}$ exhibit concave behavior with respect to $n$ (different phase flip channels); whereas when the phase flip channel is fixed (fixed $n$), $\Delta\mathcal{M}_{T,\frac{3}{4}}$ and $\Delta\mathcal{M}_{O,\frac{3}{4}}$ exhibit convex behavior with respect to $A$ (different quantum states), and $\Delta\mathcal{M}_{S,\frac{3}{4}}$ transitions from convex to concave as $A$ increases.
Finally, as shown in Figs. 2(f) and 2(i), $\Delta\mathcal{M}_{T,\frac{3}{4}}$, $\Delta\mathcal{M}_{S,\frac{3}{4}}$, and $\Delta\mathcal{M}_{O,\frac{3}{4}}$ display identical behavior in the phase flip channel as in the amplitude damping channel.

%The decay functions $\Delta\mathcal{M}_{T,\frac{3}{4}}$, $\Delta\mathcal{M}_{S,\frac{3}{4}}$ and $\Delta\mathcal{M}_{O,\frac{3}{4}}$ are shown in Fig. \ref{Fig:1} for various scenarios. Our analysis reveals that both $\Delta\mathcal{M}_{T,\frac{3}{4}}$, $\Delta\mathcal{M}_{S,\frac{3}{4}}$ and $\Delta\mathcal{M}_{O,\frac{3}{4}}$ achieve their maximum values for maximally imaginary states ($A=0$) and vanish for real states ($A=1$). For a fixed quantum state (fixed $A$), the decay functions exhibit concavity with respect to the channel parameters $m,n,p$. Conversely, for a fixed channel (fixed $m,n,p$), the decay functions are convex and decreasing with respect to $A$.

In general, to compare the rates of change of two bivariate functions, one can analyze the magnitudes of their gradient vectors at corresponding points. This provides insight into the rate of change at those points and aids in evaluating the stability of the functions. While gradient vector magnitudes are informative, visual representations are often more intuitive and easier to interpret. Plotting a three-dimensional graph of the function can effectively highlight changes in slope and curvature across different directions, which may not be easily discernible through numerical calculations. Such visualizations capture the overall trend of change and offer a clearer understanding of stability.
As observed in Fig. \ref{Fig:1}, the Sandwiched R\'{e}nyi relative entropy of imaginarity decreases rapidly (green graphs), while the Tsallis relative $\alpha$-entropy of imaginarity (red graphs) demonstrates greater stability under the quantum channel. Consequently, under the quantum channel $\epsilon$, the Tsallis relative $\alpha$-entropy of imaginarity exhibits a smaller attenuation difference, indicating that it retains more information during transmission and exhibits higher stability.

\section{Quantum state order of imaginarity measures}
In this section, we mainly study the ordering of singe-qubit pure states under the different imaginarity measures. Firstly, we discuss the ordering of quantum states based on $\mathcal{M}_{T,\alpha}$, $\mathcal{M}_{S,\alpha}$, $\mathcal{M}_{O,\alpha}$. Then we also discuss the ordering of single-qubit states after passing through the real channel under the imagimarity measures. The real channels mainly involved bit flip channel.

Let $\mathcal{M}_{A}$ and $\mathcal{M}_{B}$ denote two different measures of imaginarity. For any two single-qubit states $\rho_{1}$ and $\rho_{2}$, if the following relationship holds:
\begin{equation}
	\mathcal{M}_{A}(\rho_{1})\leq \mathcal{M}_{A}(\rho_{2})\Leftrightarrow \mathcal{M}_{B}(\rho_{1})\leq \mathcal{M}_{B}(\rho_{2}),\nonumber
\end{equation}
then the measures $\mathcal{M}_{A}$ and $\mathcal{M}_{B}$ are said to be of the same order. Conversely, if this relation is not satisfied, the measures $\mathcal{M}_{A}$ and $\mathcal{M}_{B}$ are considered to be of different orders.

For any qubit mixed state $\rho^\prime$, there exists a $2\times 2$ orthogonal matrix $O$ such that \cite{Hickey1}
\begin{eqnarray}
	\rho=O\rho^\prime O^T=\begin{pmatrix}
		\frac{1}{2} & x-iy\\
		x+iy & \frac{1}{2}
	\end{pmatrix},\label{mixed}
\end{eqnarray}
where $x$ and $y$ are non-negative real numbers such that $x^2+y^2\leq \frac{1}{4}$. $\rho^\prime$ and $\rho$ have the same imaginarity. It is important to note that if $\rho$ is a pure state, then $x^2 + y^2 = \frac{1}{4}$. Therefore, we focus exclusively on the pure state $\rho$.

Using the definitions of the measures $\mathcal{M}_{T,\alpha}$, $\mathcal{M}_{S,\alpha}$, and $\mathcal{M}_{O,\alpha}$ for qubit pure states, we can calculate:
\begin{eqnarray}
	\mathcal{M}_{T,\alpha}(\rho)&=&1-\left(x+\frac{1}{2}\right)^{\frac{1}{\alpha}},\nonumber\\
	\mathcal{M}_{S,\alpha}(\rho)&=&\frac{\left(x+\frac{1}{2}\right)^{\frac{\alpha }{1-\alpha }}-1}{\alpha -1},\nonumber\\
	\mathcal{M}_{O,\alpha}(\rho)&=&\frac{\left(x+\frac{1}{2}\right)^{\frac{1}{\alpha }-1}-1}{\alpha -1}.\nonumber
\end{eqnarray}
From the derivations of the imaginarity measures with respect to $x$,
\begin{eqnarray}
	\partial_x \mathcal{M}_{T,\alpha}(\rho)&=&-\frac{\left(x+\frac{1}{2}\right)^{\frac{1}{\alpha}-1}}{\alpha }, \nonumber\\
	\partial_x \mathcal{M}_{S,\alpha}(\rho)&=&-\frac{2 \alpha  \left(x+\frac{1}{2}\right)^{\frac{\alpha }{1-\alpha }}}{(\alpha -1)^2 (2 x+1)},\nonumber\\
	\partial_x \mathcal{M}_{O,\alpha}(\rho)&=&-\frac{4 \left(x+\frac{1}{2}\right)^{\frac{1}{\alpha}}}{\alpha(2 x+1)^2}.\nonumber
\end{eqnarray}
Since $x$ is a non-negative real number, and given that $\frac{1}{\alpha} \in (1,2]$ and $\frac{\alpha}{1-\alpha} \geq 1$, it follows that $X^{\frac{1}{\alpha}} \geq 0$, $X^{\frac{1}{\alpha}-1} \geq 0$, and $X^{\frac{\alpha}{1-\alpha}} \geq 0$, where $X = x + \frac{1}{2}$. 
Consequently, we conclude that $\partial_x \mathcal{M}_{T,\alpha}(\rho) \leq 0$, $\partial_x \mathcal{M}_{S,\alpha}(\rho) \leq 0$, and $\partial_x \mathcal{M}_{O,\alpha}(\rho) \leq 0$. This indicates that $\mathcal{M}_{T,\alpha}(\rho)$, $\mathcal{M}_{S,\alpha}(\rho)$, and $\mathcal{M}_{O,\alpha}(\rho)$ are all decreasing functions.

Based on the analysis above, we observe that $\mathcal{M}_{T,\alpha}$, $\mathcal{M}_{S,\alpha}$, and $\mathcal{M}_{O,\alpha}$ exhibit the same monotonicity for single-qubit states. Therefore, we can draw the following conclusions.
\begin{proposition}
	For any two single-qubit states $\rho_{1}$ and $\rho_{2}$ in the form given by formula (\ref{mixed}), the imaginarity measures $\mathcal{M}_{T,\alpha}$, $\mathcal{M}_{S,\alpha}$, and $\mathcal{M}_{O,\alpha}$ satisfy the following relations:
	\begin{eqnarray}
		&&\mathcal{M}_{T,\alpha}(\rho_{1})\geq \mathcal{M}_{T,\alpha}(\rho_{2}) \Leftrightarrow \mathcal{M}_{S,\alpha}(\rho_{1})\geq \mathcal{M}_{S,\alpha}(\rho_{2})\nonumber\\ 
		&&\Leftrightarrow \mathcal{M}_{O,\alpha}(\rho_{1})\geq \mathcal{M}_{O,\alpha}(\rho_{2}),\nonumber
	\end{eqnarray}
	where $\alpha \in [\frac{1}{2},1)$.
\end{proposition}

It is well known that the quantum channel can change the quantum state, furthermore it can affect the quantum state order also. Next, we study whether the order of quantum states will change after they pass through a quantum channel under the given imaginarity measure and quantum channel.
For an imaginarity measure $\mathcal{M}$ and quantum channel $\varepsilon$, for two quantum pure states $\rho_1$ and $\rho_2$ \cite{chen2023},
\begin{eqnarray}
	\mathcal{M}(\rho_1)\geq \mathcal{M}( \rho_2) \Leftrightarrow \mathcal{M}(\varepsilon(\rho_1))\geq \mathcal{M}(\varepsilon(\rho_2))
\end{eqnarray}
or
\begin{eqnarray}
	\mathcal{M}(\rho_1)\leq \mathcal{M}(\rho_2) \Leftrightarrow \mathcal{M}(\varepsilon(\rho_1))\leq \mathcal{M}(\varepsilon(\rho_2)).
\end{eqnarray}
If the above relations hold, we say that the quantum channel $\varepsilon$ does not change the order of the quantum states; if not, it indicates that the quantum channel $\varepsilon$ alters the order of the quantum states. In this section, we primarily focus on the order of quantum states based on imaginarity measures after they pass through the bit flip channel.

%We first examine the case of the bit flip channel $\varepsilon_{BF}$ in conjunction with the imaginarity measures $\mathcal{M}_{T,\alpha}$, $\mathcal{M}_{S,\alpha}$, and $\mathcal{M}_{O,\alpha}$.
The bit flip channel $\varepsilon_{BF}$ is represented by the Kraus operators $\{K_{0} = \sqrt{m}\mathbb{I}, K_{1} = \sqrt{1-m}\sigma_{x}\}$, where $\sigma_x$ is the Pauli operator. The qubit pure state $\rho$ is expressed as in Eq. (\ref{mixed}) with the condition $x^2 + y^2 = \frac{1}{4}$.
\begin{proposition}\label{BForder}
	The order of the quantum states characterized by $\mathcal{M}_{T,\alpha}$, $\mathcal{M}_{S,\alpha}$, and $\mathcal{M}_{O,\alpha}$ remains unchanged after a qubit pure state passes through the bit flip channel. The following relationships hold between these measures:
	\begin{eqnarray}
		&&\mathcal{M}_{T,\alpha}(\rho_{1})\geq \mathcal{M}_{T,\alpha}(\rho_{2})\nonumber\\
		&&\Leftrightarrow \mathcal{M}_{T,\alpha}(\varepsilon_{BF}(\rho_{1}))\geq \mathcal{M}_{T,\alpha}(\varepsilon_{BF}(\rho_{2})),\nonumber\\
		&&\mathcal{M}_{S,\alpha}(\rho_{1})\geq \mathcal{M}_{S,\alpha}(\rho_{2})\nonumber\\
		&&\Leftrightarrow \mathcal{M}_{S,\alpha}(\varepsilon_{BF}(\rho_{1}))\geq \mathcal{M}_{S,\alpha}(\varepsilon_{BF}(\rho_{2})),\nonumber\\
		&&\mathcal{M}_{O,\alpha}(\rho_{1})\geq \mathcal{M}_{O,\alpha}(\rho_{2})\nonumber\\
		&&\Leftrightarrow \mathcal{M}_{O,\alpha}(\varepsilon_{BF}(\rho_{1}))\geq \mathcal{M}_{O,\alpha}(\varepsilon_{BF}(\rho_{2})).\nonumber
	\end{eqnarray}
	where $\alpha \in [\frac{1}{2},1)$.
\end{proposition}

{\sf [Proof]} 
The pure state of the qubit system after passing through the bit flip channel $\varepsilon_{BF}$ is given by
\begin{equation}
	\begin{aligned}
		\varepsilon_{BF}(\rho)&=K_{0}\rho K_{0}^{\dagger}+K_{1}\rho K_{1}^{\dagger}\\
		&=\begin{pmatrix}
			\frac{1}{2} & x+i (1-2 m) y \\
			x+i (2 m-1) y & \frac{1}{2} \\
		\end{pmatrix},
	\end{aligned}
\end{equation}
where $\rho$ is expressed by Eq. (\ref{mixed}) and $x^2 + y^2 = \frac{1}{4}$.

It is easy to derive that
\begin{align}
	\mathcal{M}_{T,\alpha}(\varepsilon_{BF}(\rho))&=1-2^{-\frac{1}{\alpha}} \Big[\frac{x}{\sqrt{(1-2 m)^2 y^2+x^2}}+1\Big]^{\frac{1}{\alpha}},\nonumber\\
	\mathcal{M}_{S,\alpha}(\varepsilon_{BF}(\rho))&=\frac{2^{\frac{\alpha }{\alpha -1}}\Big[\frac{x}{\sqrt{(1-2 m)^2 y^2+x^2}}+1\Big]^{\frac{\alpha }{1-\alpha }}-1}{\alpha -1},\nonumber\\
	\mathcal{M}_{O,\alpha}(\varepsilon_{BF}(\rho))&=-\frac{2^{-\frac{1}{\alpha}} \Big[X_{11}[2^{\frac{1}{\alpha}}+2 (\frac{x}{X_{11}}+1)^{\frac{1}{\alpha}}]-2^{\frac{1}{\alpha}} x\Big]}{(\alpha -1)(X_{11}+x)},\nonumber
\end{align}
where $X_{11}=\sqrt{(1-2 m)^2 y^2+x^2}$.

From the derivations of the imaginarity measures with respect to $x$, we have
\begin{align}
	\partial_x \mathcal{M}_{T,\alpha}(\varepsilon_{BF}(\rho))&=\frac{2^{-\frac{1}{\alpha}} (x-X_{11})\Big(\frac{x}{X_{11}}+1\Big)^{\frac{1}{\alpha}}}{\alpha\Big[(1-2 m)^2 y^2+x^2\Big]}, \nonumber\\
	\partial_x \mathcal{M}_{S,\alpha}(\varepsilon_{BF}(\rho))&=-\frac{2^{\frac{\alpha }{\alpha -1}} \alpha (X_{11}-x)\Big(\frac{x}{X_{11}}+1\Big)^{\frac{\alpha }{1-\alpha }}}{(\alpha -1)^2\Big[(1-2 m)^2 y^2+x^2\Big]},\nonumber\\
	\partial_x \mathcal{M}_{O,\alpha}(\varepsilon_{BF}(\rho))&=\frac{2^{\frac{\alpha -1}{\alpha }}(x-X_{11})\Big(\frac{x}{X_{11}}+1\Big)^{\frac{1}{\alpha}}}{\alpha \Big[x (X_{11}+x)+(1-2 m)^2 y^2\Big]}.\nonumber
\end{align}
Since $x$ and $y$ are non-negative real numbers and $x^2 + y^2 = \frac{1}{4}$, $X_{11}=\sqrt{(1-2 m)^2 y^2+x^2}$, we have $x\leq X_{11}$. Obviously, $\partial_x \mathcal{M}_{T,\alpha}(\varepsilon_{BF}(\rho))\leq 0$, $\partial_x \mathcal{M}_{S,\alpha}(\varepsilon_{BF}(\rho))\leq 0$ and $\partial_x \mathcal{M}_{O,\alpha}(\varepsilon_{BF}(\rho))\leq0$.
So, $\mathcal{M}_{T,\alpha}$, $\mathcal{M}_{S,\alpha}$ and $\mathcal{M}_{O,\alpha}$ are all decreasing functions. Thus, Proposition \ref{BForder} is true.
$\hfill\blacksquare$

\section{Discussion}
Our study demonstrates that the introduction of imaginarity measures based on Tsallis relative $\alpha$-entropy, Sandwiched R\'{e}nyi relative entropy, and Tsallis relative operator entropy provides a robust framework for quantifying the imaginary components of quantum states. These measures not only satisfy the fundamental properties required of any imaginarity measure but also exhibit distinct behaviors under various quantum channels. The results indicate that while all three measures capture the imaginarity of quantum states, they differ in their sensitivity to quantum noise.

In particular, our analysis of the decay rates of these imaginarity measures under bit flip, phase damping, and amplitude damping channels reveals that the Tsallis relative $\alpha$-entropy of imaginarity exhibits greater stability compared to the other two measures. This observation suggests that the Tsallis relative $\alpha$-entropy of imaginarity may be more robust against decoherence, making it a promising candidate for applications in quantum information processing where stability is crucial. Further research is needed to explore the practical implications of these findings and to investigate the performance of these measures in more complex quantum systems and channels. Our work paves the way for a deeper understanding of the role of imaginary numbers in quantum mechanics and their operational significance in quantum technology.

\section*{ACKNOWLEDGMENTS}
This work was supported by the National Natural Science Foundation of China (Grant Nos. 12075159 and 12175147); and Zhejiang Provincial Natural Science Foundation of China under Grant No. LZ24A050005; Science and Technology Research Project of Jiangxi Education Department No.GJJ2400606; the specific research fund of the Innovation Platform for Academicians of Hainan Province.

\section*{Appendix}
\subsection{proof of Theorem \ref{main1}}
We prove that (\rmnum{1}) to (\rmnum{3}) satisfies (M1), (M2) and (M5).

(\rmnum{1}) We first show that $\mathcal{M}_{T,\alpha}(\rho)$ satisfies (M1). When $\alpha \in [\frac{1}{2},1)$, it follows that $\text{tr}(\rho ^{\alpha}\sigma^{1-\alpha}) - 1 \leq 0$, with equality if and only if $\rho = \sigma$ \cite{Rastegin}. This implies that $\{\text{tr}(\rho ^{\alpha}\sigma^{1-\alpha})\}^{\frac{1}{\alpha}} \leq 1$ if and only if $\rho = \sigma$. Consequently, $\mathcal{M}_{T,\alpha}$ satisfies (M1).

Next, we establish that $D_{\alpha}(\rho || \sigma)$ is nonincreasing under a completely positive trace-preserving (CPTP) map $\varepsilon$ when $\alpha \in [\frac{1}{2},1)$, such that
\begin{eqnarray}
	D_{\alpha}[\varepsilon(\rho)||\varepsilon(\sigma)]\leq D_{\alpha}(\rho ||\sigma)
\end{eqnarray}
for any states $\rho$ and $\sigma$. Therefore, we obtain
\begin{eqnarray}
	\text{tr}(\rho^{\alpha}\sigma^{1-\alpha})\leq \text{tr}(\varepsilon(\rho)^{\alpha}\varepsilon(\sigma)^{1-\alpha}),
\end{eqnarray}
so
\begin{eqnarray}\label{TTT6}
	\{\text{tr}(\rho^{\alpha}\sigma^{1-\alpha})\}^{\frac{1}{\alpha}}\leq \{\text{tr}(\varepsilon(\rho)^{\alpha}\varepsilon(\sigma)^{1-\alpha})\}^{\frac{1}{\alpha}}
\end{eqnarray}
Let $\varepsilon$ be a real operation within CPTP mappings. Suppose $\sigma_* \in \mathcal{F}$; then we have
\begin{eqnarray}
	&&\min_{\sigma \in \mathcal{F}} \Big\{1-\Big[\text{tr}(\rho^{\alpha}\sigma^{1-\alpha})\Big]^{\frac{1}{\alpha}}\Big\}\nonumber\\
	&&=\Big\{1-\Big[\text{tr}(\rho^{\alpha}\sigma_*^{1-\alpha})\Big]^{\frac{1}{\alpha}}\Big\}\nonumber\\
	&&\geq\Big\{1-\Big[\text{tr}(\varepsilon(\rho)^{\alpha}\varepsilon(\sigma_*)^{1-\alpha})\Big]^{\frac{1}{\alpha}}\Big\}\nonumber\\
	&&\geq\min_{\sigma \in \mathcal{F}} \Big\{1-\Big[\text{tr}(\varepsilon(\rho)^{\alpha}\sigma^{1-\alpha})\Big]^{\frac{1}{\alpha}}\Big\}\nonumber\\
	&&=\mathcal{M}_{T,\alpha}(\varepsilon(\rho)),\nonumber
\end{eqnarray}
where the first inequality follows from (\ref{TTT6}), and the second inequality holds because $\varepsilon(\sigma_*) \in \mathcal{F}$. Thus, we conclude that $\mathcal{M}_{T,\alpha}(\rho)$ satisfies (M2).

Now we prove that $\mathcal{M}_{T,\alpha}(\rho)$ also satisfies (M5). For the direct sum state $\rho = p_1\rho_1 \oplus p_2\rho_2$ and real direct sum state $\sigma = q_1\sigma_1 \oplus \sigma_2 q_2$, where $p_1, p_2, q_1, q_2 \in (0,1)$, $p_1^2 + p_2^2 = 1$, and $q_1^2 + q_2^2 = 1$, we have
\begin{align}
	&\max_{\sigma \in \mathcal{F}}\Big[\text{tr}(\rho^{\alpha}\sigma^{1-\alpha})\Big]\nonumber\\
	&=\max_{\sigma \in \mathcal{F}}\Big[\text{tr}\Big((p_1\rho _{1}\oplus p_2\rho _{2})^{\alpha}(q_1\sigma_{1}\oplus \sigma q_2)^{1-\alpha}\Big)\Big]\nonumber\\
	&=\max_{q_1,q_2}\Big\{(p_1^\alpha q_1^{1-\alpha})\max_{\sigma_{1}}\text{tr}(\rho_1^{\alpha}\sigma_1^{1-\alpha})\nonumber\\
	&~+(p_2^\alpha q_2^{1-\alpha})\max_{\sigma_{2}}\text{tr}(\rho_2^{\alpha}\sigma_2^{1-\alpha}) \Big\}\nonumber\\
	&=\max_{q_1,q_2}\Big\{p_1^\alpha q_1^{1-\alpha} t_1+p_2^\alpha q_2^{1-\alpha} t_2 \Big\}\nonumber\\
	&=\Big[p_1t_1^{\frac{1}{\alpha}}+p_2t_2^{\frac{1}{\alpha}}\Big]^{\alpha},\nonumber
\end{align}
where $t_1=\max_{\sigma_{1}}\text{tr}(\rho_1^{\alpha}\sigma_1^{1-\alpha})$, $t_2=\max_{\sigma_{2}}\text{tr}(\rho_2^{\alpha}\sigma_2^{1-\alpha})$.
The last equality follows from H\"{o}lder inequality:
\begin{align}
	&p_1^\alpha q_1^{1-\alpha} t_1+p_2^\alpha q_2^{1-\alpha} t_2\nonumber\\
	&\leq \Big[(q_1^{1-\alpha})^{\frac{1}{1-\alpha}}+(q_2^{1-\alpha})^{\frac{1}{1-\alpha}}\Big]^{1-\alpha}\Big[(p_1^{\alpha}t_1)^{\frac{1}{\alpha}}+(p_2^{\alpha}t_2)^{\frac{1}{\alpha}}\Big]^{\alpha}\nonumber\\
	&=\Big[p_1 t_1^{\frac{1}{\alpha}}+p_2 t_2^{\frac{1}{\alpha}}\Big]^{\alpha},\nonumber
\end{align}
where equality holds when $\frac{q_1}{q_2} = \frac{p_1 t_1^{\frac{1}{\alpha}}}{p_2 t_2^{\frac{1}{\alpha}}}$.
Consequently,
\begin{align}
	\max_{\sigma \in \mathcal{F}}\Big\{\Big[\text{tr}(\rho^{\alpha}\sigma^{1-\alpha})\Big]^{\frac{1}{\alpha}}\Big\}&=\Big\{\max_{\sigma \in \mathcal{F}}\text{tr}(\rho^{\alpha}\sigma^{1-\alpha})\Big\}^{\frac{1}{\alpha}}\nonumber\\
	&=p_1 t_1^{\frac{1}{\alpha}}+p_2 t_2^{\frac{1}{\alpha}}.\nonumber
\end{align}
We have thus demonstrated that $\mathcal{M}_{F,\alpha}(\rho)$ satisfies (M5).

(\rmnum{2}) It is shown that for $\alpha \in [\frac{1}{2}, 1)$, $F_{\alpha}(\sigma || \rho) \geq 0$, with equality holding if and only if $\sigma = \rho$ \cite{Muller37}. This is equivalent to
\begin{eqnarray}
	\text{tr}[(\rho^{\frac{1-\alpha }{2\alpha }}\sigma \rho ^{\frac{1-\alpha }{2\alpha }})^{\alpha }]\leq 1 \nonumber
\end{eqnarray}
with equality holding if and only if $\sigma = \rho$, and further equivalent to
\begin{eqnarray}
	\{\text{tr}[(\rho^{\frac{1-\alpha }{2\alpha }}\sigma \rho ^{\frac{1-\alpha }{2\alpha }})^{\alpha }]\}^{\frac{1}{1-\alpha}}\leq 1 \nonumber
\end{eqnarray}
with equality holding if and only if $\sigma = \rho$.
Thus, $\mathcal{M}_{S,\alpha}(\rho)$ satisfies (M1).

For $\alpha \in [\frac{1}{2}, 1)$, it has been shown that for any states $\rho, \sigma$ and any CPTP map $\varepsilon$ \cite{Muller37},
\begin{eqnarray}
	F_{\alpha}(\varepsilon(\sigma)||\varepsilon(\rho ))\leq F_{\alpha }(\sigma||\rho). \nonumber
\end{eqnarray}
This implies
\begin{eqnarray}
	\text{tr}\Big[\Big(\varepsilon(\rho)^{\frac{1-\alpha }{2\alpha }}\varepsilon(\sigma )\varepsilon(\rho )^{\frac{1-\alpha }{2\alpha }}\Big)^{\alpha }\Big] \geq \text{tr}[(\rho ^{\frac{1-\alpha }{2\alpha }}\sigma \rho^{\frac{1-\alpha }{2\alpha }})^{\alpha }], \nonumber
\end{eqnarray}
and
\begin{eqnarray}
	&&\Big\{\text{tr}\Big[\Big(\varepsilon(\rho)^{\frac{1-\alpha }{2\alpha }}\varepsilon(\sigma )\varepsilon(\rho )^{\frac{1-\alpha }{2\alpha }}\Big)^{\alpha }\Big]\Big\}^{\frac{1}{1-\alpha}}\nonumber\\
	&&\geq \Big\{\text{tr}\Big[(\rho ^{\frac{1-\alpha }{2\alpha }}\sigma \rho^{\frac{1-\alpha }{2\alpha }})^{\alpha }\Big]\Big\}^{\frac{1}{1-\alpha}}. \nonumber
\end{eqnarray}
For any real operation $\varepsilon$ within CPTP mappings, suppose $\sigma_{*} \in \mathcal{F}$; then we obtain
\begin{align}
	&\min_{\sigma \in \mathcal{F}}\frac{1}{\alpha-1}\Bigg\{\Big\{\text{tr}\Big[(\rho^{\frac{1-\alpha}{2\alpha}}\sigma\rho^{\frac{1-\alpha}{2\alpha}})^{\alpha}\Big]\Big\}^{\frac{1}{1-\alpha}}-1\Bigg\}\nonumber\\
	&=\frac{1}{\alpha-1}\Bigg\{\Big\{\text{tr}\Big[(\rho^{\frac{1-\alpha}{2\alpha}}\sigma_{*}\rho^{\frac{1-\alpha}{2\alpha}})^{\alpha}\Big]\Big\}^{\frac{1}{1-\alpha}}-1\Bigg\} \nonumber \\
	&\geq\frac{1}{\alpha-1}\Bigg\{\Big\{\text{tr}\Big[\Big(\varepsilon(\rho)^{\frac{1-\alpha }{2\alpha }}\varepsilon(\sigma_{*})\varepsilon(\rho)^{\frac{1-\alpha }{2\alpha }}\Big)^{\alpha }\Big]\Big\}^{\frac{1}{1-\alpha}}-1\Bigg\} \nonumber \\
	&\geq\min_{\sigma \in \mathcal{F}}\frac{1}{\alpha-1}\Bigg\{\Big\{\text{tr}\Big[\Big(\varepsilon (\rho)^{\frac{1-\alpha }{2\alpha }}\sigma \varepsilon(\rho)^{\frac{1-\alpha }{2\alpha }}\Big)^{\alpha }\Big]\Big\}^{\frac{1}{1-\alpha}}-1\Bigg\} \nonumber\\
	&=\mathcal{M}_{S,\alpha}(\varepsilon(\rho)),\nonumber
\end{align}
where the last inequality follows from $\varepsilon(\sigma_{*}) \in \mathcal{F}$. This proves that $\mathcal{M}_{S,\alpha}(\rho)$ satisfies (M2).

To prove (M5), let us s consider again $\rho=p_1\rho _{1}\oplus p_2\rho _{2}$ and $\sigma =q_1\sigma_{1}\oplus \sigma_2 q_2$, we have
\begin{align}
	&\max_{\sigma \in \mathcal{F}}\Big\{\text{tr}\Big[(\rho^{\frac{1-\alpha}{2\alpha}}\sigma\rho^{\frac{1-\alpha}{2\alpha}})^{\alpha}\Big]\Big\}\nonumber\\
	&=\max_{\sigma \in \mathcal{F}}\Big\{\text{tr}\Big[\Big((p_1\rho _{1}\oplus p_2\rho _{2})^{\frac{1-\alpha}{2\alpha}}(q_1\sigma_{1}\oplus \sigma_2 q_2)\cdot\nonumber\\
	&~~~~(p_1\rho _{1}\oplus p_2\rho _{2})^{\frac{1-\alpha}{2\alpha}}\Big)^{\alpha}\Big]\Big\}\nonumber\\
	&=\max_{q_1,q_2}\Big\{(p^{1-\alpha}_1q_1^{\alpha})\max_{\sigma_{1}}\text{tr}\Big[(\rho_1^{\frac{1-\alpha}{2\alpha}}\sigma_1 \rho_1^{\frac{1-\alpha}{2\alpha}})^{\alpha}\Big]\nonumber\\
	&~+(p^{1-\alpha}_2q_2^{\alpha})\max_{\sigma_{2}}\text{tr}\Big[(\rho_2^{\frac{1-\alpha}{2\alpha}}\sigma_2 \rho_2^{\frac{1-\alpha}{2\alpha}})^{\alpha}\Big]\Big\}\nonumber\\
	&=\max_{q_1,q_2}\Big\{(p^{1-\alpha}_1q_1^{\alpha})t^{'}_1+(p^{1-\alpha}_2q_2^{\alpha})t^{'}_2\Big\}\nonumber\\
	&=[p_1 (t^{'}_1)^{\frac{1}{1-\alpha}}+p_2 (t^{'}_2)^{\frac{1}{1-\alpha}}]^{1-\alpha},\nonumber
\end{align}
where $t^{'}_1=\max_{\sigma_{1}}\text{tr}\Big[(\rho_1^{\frac{1-\alpha}{2\alpha}}\sigma_1 \rho_1^{\frac{1-\alpha}{2\alpha}})^{\alpha}\Big]$, $t^{'}_2=\max_{\sigma_{2}}\text{tr}\Big[(\rho_2^{\frac{1-\alpha}{2\alpha}}\sigma_2 \rho_2^{\frac{1-\alpha}{2\alpha}})^{\alpha}\Big]$.
The last equality is used the H\"{o}lder inequality such that
\begin{align}
	&q^{\alpha}_1(p_1^{1-\alpha}t^{'}_1)+q^{\alpha}_2(p_2^{1-\alpha}t^{'}_2)\nonumber\\
	&\leq \Big[(q_1^{\alpha})^{\frac{1}{\alpha}}+(q_2^{\alpha})^{\frac{1}{\alpha}}\Big]^\alpha\Big[(p_1^{\alpha}t^{'}_1)^{\frac{1}{1-\alpha}}+(p_2^{\alpha}t^{'}_2)^{\frac{1}{1-\alpha}}\Big]^{1-\alpha}\nonumber\\
	&=\Big[p_1 (t^{'}_1)^{\frac{1}{1-\alpha}}+p_2 (t^{'}_2)^{\frac{1}{1-\alpha}}\Big]^{1-\alpha},\nonumber
\end{align}
the equality holds when $\frac{q_1}{q_2}=\frac{p_1 (t^{'}_1)^{\frac{1}{1-\alpha}}}{p_2 (t^{'}_2)^{\frac{1}{1-\alpha}}}$.
Consequently,
\begin{align}
	&\max_{\sigma \in \mathcal{F}}\Big\{\Big[\text{tr}\Big((\rho^{\frac{1-\alpha}{2\alpha}}\sigma\rho^{\frac{1-\alpha}{2\alpha}})^{\alpha}\Big)\Big]^{\frac{1}{1-\alpha}}\Big\}\nonumber\\
	&=\Big\{\max_{\sigma \in \mathcal{F}}\text{tr}\Big[(\rho^{\frac{1-\alpha}{2\alpha}}\sigma\rho^{\frac{1-\alpha}{2\alpha}})^{\alpha}\Big]\Big\}^{\frac{1}{1-\alpha}}\nonumber\\
	&=p_1 (t^{'}_1)^{\frac{1}{1-\alpha}}+p_2 (t^{'}_2)^{\frac{1}{1-\alpha}}\nonumber
\end{align}
We then proved that $\mathcal{M}_{S,\alpha}(\rho)$ satisfies (M5).

(\rmnum{3})
It is easy to find that $\mathcal{M}_{O,\alpha}(\rho)=0$, iff $\rho$ is an real state. It follows that $\mathcal{M}_{O,\alpha}(\rho)$ satisfies (M1).
For $\alpha \in [\frac{1}{2},1),$ it has been shown that for any state $\sigma$, and $\rho$ and any CPTP map $\varepsilon$\cite{Guo2020},
\begin{align}
	&\mathrm{tr}\Big[\rho^{\frac{1}{2}}(\rho^{-\frac{1}{2}}\sigma \rho^{-\frac{1}{2}})^{1-\alpha}\rho^{\frac{1}{2}}\Big] \nonumber\\
	&\leq \mathrm{tr}\Big[\varepsilon(\rho)^{\frac{1}{2}}\Big(\varepsilon(\rho)^{-\frac{1}{2}}\varepsilon(\sigma) \varepsilon(\rho)^{-\frac{1}{2}}\Big)^{1-\alpha}\varepsilon(\rho)^{\frac{1}{2}}\Big].\nonumber
\end{align}
It means
\begin{align}
	&\Big\{\text{tr}\Big[\rho^{\frac{1}{2}}(\rho^{-\frac{1}{2}}\sigma \rho^{-\frac{1}{2}})^{1-\alpha}\rho^{\frac{1}{2}}\Big]\Big\}^{\frac{1}{\alpha}} \nonumber\\
	&\leq \Big\{\text{tr}\Big[\varepsilon(\rho)^{\frac{1}{2}}\Big(\varepsilon(\rho)^{-\frac{1}{2}}\varepsilon(\sigma) \varepsilon(\rho)^{-\frac{1}{2}}\Big)^{1-\alpha}\varepsilon(\rho)^{\frac{1}{2}}\Big]\Big\}^{\frac{1}{\alpha}}. \nonumber
\end{align}
For any real operation $\varepsilon$ within CPTP mappings, suppose $\sigma_{*} \in \mathcal{F}$, we get
\begin{align}
	&\min_{\sigma \in \mathcal{F}}\frac{1}{\alpha-1} \Bigg\{\Big[\text{tr}\Big(\rho^{\frac{1}{2}}(\rho^{-\frac{1}{2}}\sigma \rho^{-\frac{1}{2}})^{1-\alpha}\rho^{\frac{1}{2}}\Big)\Big]^{\frac{1}{\alpha}}-1\Bigg\}\nonumber\\
	&=\frac{1}{\alpha-1}\Bigg\{\Big[\text{tr}\Big(\rho^{\frac{1}{2}}(\rho^{-\frac{1}{2}}\sigma_{*} \rho^{-\frac{1}{2}})^{1-\alpha}\rho^{\frac{1}{2}}\Big)\Big]^{\frac{1}{\alpha}}-1\Bigg\}\nonumber\\
	&\geq \frac{1}{\alpha-1}\Bigg\{\Bigg[ \mathrm{tr}\Big[\varepsilon(\rho)^{\frac{1}{2}}\Big(\varepsilon(\rho)^{-\frac{1}{2}}\varepsilon(\sigma_{*}) \varepsilon(\rho)^{-\frac{1}{2}}\Big)^{1-\alpha}\varepsilon(\rho)^{\frac{1}{2}}\Big]\Bigg]^{\frac{1}{\alpha}}\nonumber\\
	&~~-1\Bigg\}\nonumber\\
	&\geq \min_{\sigma \in \mathcal{F}}\frac{1}{\alpha-1} \Bigg\{\Bigg[ \mathrm{tr}\Big[\varepsilon(\rho)^{\frac{1}{2}}\Big(\varepsilon(\rho)^{-\frac{1}{2}}\sigma \varepsilon(\rho)^{-\frac{1}{2}}\Big)^{1-\alpha}\varepsilon(\rho)^{\frac{1}{2}}\Big]\Bigg]^{\frac{1}{\alpha}}\nonumber\\
	&~~-1\Bigg\}\nonumber\\
	&=\mathcal{M}_{O,\alpha}(\varepsilon(\rho)),\nonumber
\end{align}
where the second inequality is because of $\varepsilon(\sigma_{*})\in\mathcal{F}$. This proves that $\mathcal{M}_{O,\alpha}(\rho)$ satisfies (M2).

Then we prove that $\mathcal{M}_{O,\alpha}(\rho)$ satisfies (M5). Let us consider again $\rho=p_1\rho _{1}\oplus p_2\rho _{2}$ and $\sigma =q_1\sigma_{1}\oplus \sigma_2 q_2$, we have
\begin{align}
	&\max_{\sigma \in \mathcal{F}}\Bigg\{\text{tr}\left[\rho^{\frac{1}{2}}\left(\rho^{-\frac{1}{2}}\sigma \rho^{-\frac{1}{2}}\right)^{1-\alpha}\rho^{\frac{1}{2}}\right]\Bigg\}\nonumber\\
	&=\min_{\sigma \in \mathcal{F}} \Bigg\{\text{tr}\Big[(p_1\rho _{1}\oplus p_2\rho _{2})^{\frac{1}{2}}\Big((p_1\rho _{1}\oplus p_2\rho _{2})^{-\frac{1}{2}}(q_1\sigma_{1}\oplus \sigma_2 q_2)\cdot \nonumber\\
	&~~~~(p_1\rho _{1}\oplus p_2\rho _{2})^{-\frac{1}{2}}\Big)^{1-\alpha}(p_1\rho _{1}\oplus p_2\rho _{2})^{\frac{1}{2}}\Big] \Bigg\}\nonumber\\
	&=\max_{\sigma \in \mathcal{I}}\Big\{\mathrm{tr}\Big[(p^{\frac{1}{2}}_1\rho^{\frac{1}{2}}_1 \oplus p^{\frac{1}{2}}_2\rho^{\frac{1}{2}}_2) (p^{-\frac{1}{2}}_1 q_1 p^{-\frac{1}{2}}_1 \rho^{-\frac{1}{2}}_1\sigma_1 \rho^{-\frac{1}{2}}_1 \oplus \nonumber\\
	&~~~~p^{-\frac{1}{2}}_2 q_2 p^{-\frac{1}{2}}_2 \rho^{-\frac{1}{2}}_2 \sigma_2 \rho^{-\frac{1}{2}}_2)^{1-\alpha} (p^{\frac{1}{2}}_1\rho^{\frac{1}{2}}_1 \oplus p^{\frac{1}{2}}_2\rho^{\frac{1}{2}}_2)\Big]\Big\}\nonumber\\
	&=\max_{q_1,q_2}\Big\{\mathrm{tr}\Big[p^{\alpha}_1 q^{1-\alpha}_1 \rho^{\frac{1}{2}}_1(\rho^{-\frac{1}{2}}_1\sigma_1 \rho^{-\frac{1}{2}}_1)^{1-\alpha}\rho^{\frac{1}{2}}_1\nonumber\\
	&~~+p^{\alpha}_2 q^{1-\alpha}_2 \rho^{\frac{1}{2}}_2 (\rho^{-\frac{1}{2}}_2\sigma_2 \rho^{-\frac{1}{2}}_2)^{1-\alpha}\rho^{\frac{1}{2}}_2\Big]\Big\}\nonumber\\
	&=\max_{q_1,q_2}\Big[p^{\alpha}_1 q^{1-\alpha}_1 t^{''}_1 +p^{\alpha}_2 q^{1-\alpha}_2 t^{''}_2\Big],\nonumber\\
	&=\Big[p_1 (t^{''}_1)^{\frac{1}{\alpha}}+p_2 (t^{''}_2)^{\frac{1}{\alpha}}\Big]^{\alpha},\nonumber
\end{align}
where $t^{''}_1=\max_{\sigma_1} \Big\{\mathrm{tr}\Big[\rho^{\frac{1}{2}}_1\left(\rho^{-\frac{1}{2}}_1\sigma_1\rho^{-\frac{1}{2}}_1\right)^{1-\alpha}\rho^{\frac{1}{2}}_1\Big]\Big\}$,
$t^{''}_2=\max_{\sigma_2} \Big\{\mathrm{tr}\Big[\rho^{\frac{1}{2}}_2\left(\rho^{-\frac{1}{2}}_2\sigma_2 \rho^{-\frac{1}{2}}_2\right)^{1-\alpha}\rho^{\frac{1}{2}}_2\Big]\Big\}$,
the last equality is used the H\"{o}lder inequality such that
\begin{align}
	&p_1^\alpha q_1^{1-\alpha} t^{''}_1+p_2^\alpha q_2^{1-\alpha} t^{''}_2\nonumber\\
	&\leq \Big[(q_1^{1-\alpha})^{\frac{1}{1-\alpha}}+ (q_2^{1-\alpha})^{\frac{1}{1-\alpha}}\Big]^{1-\alpha}\Big[(p_1^{\alpha}t^{''}_1)^{\frac{1}{\alpha}}+(p_2^{\alpha}t^{''}_2)^{\frac{1}{\alpha}}\Big]^{\alpha}\nonumber\\
	&=\Big[p_1 (t^{''}_1)^{\frac{1}{\alpha}}+p_2 (t^{''}_2)^{\frac{1}{\alpha}}\Big]^{\alpha},\nonumber
\end{align}
the equality holds when $\frac{q_1}{q_2}=\frac{p_1 (t^{''}_1)^{\frac{1}{\alpha}}}{p_2 (t^{''}_2)^{\frac{1}{\alpha}}}$.
Consequently,
\begin{align}
	&\max_{\sigma \in \mathcal{F}}\Bigg\{\Big[\text{tr}\Big(\rho^{\frac{1}{2}}(\rho^{-\frac{1}{2}}\sigma \rho^{-\frac{1}{2}})^{1-\alpha}\rho^{\frac{1}{2}}\Big)\Big]^{\frac{1}{\alpha}}\Bigg\}\nonumber\\
	&=\Big\{\max_{\sigma \in \mathcal{F}}\text{tr}\Big[\rho^{\frac{1}{2}}(\rho^{-\frac{1}{2}}\sigma \rho^{-\frac{1}{2}})^{1-\alpha}\rho^{\frac{1}{2}}\Big]\Big\}^{\frac{1}{\alpha}}\nonumber\\
	&=p_1 t_1^{\frac{1}{\alpha}}+p_2 t_2^{\frac{1}{\alpha}}.\nonumber
\end{align}
We then proved that $\mathcal{M}_{O,\alpha}(\rho)$ satisfies (M5).

\end{document}